\documentclass[aps,preprint]{revtex4-2}

\usepackage[utf8]{inputenc}
\usepackage[T1]{fontenc}
\usepackage{amsmath}
\usepackage{amsthm}
\usepackage{amssymb}
\usepackage{hyperref}
\usepackage{dsfont}
\usepackage{graphicx}
\usepackage{tikz-cd}

\hypersetup{colorlinks=true, linkcolor=blue, citecolor=blue}

\newcommand{\Real}{\mathbb{R}}

\newcommand{\Comp}{\mathbb{C}}
\newcommand{\seq}{\subseteq}

\newcommand{\pspace}{\Real^3 \setminus \{0\}}

\newcommand{\Lightcone}{\mathcal{L}_+}

\newcommand{\poincareGroup}{\mathrm{ISO}^+(3,1)}
\newcommand{\lorentzGroup}{\mathrm{SO}^+(3,1)}
\newcommand{\Poincare}{Poincar\'{e} }
\newcommand{\SO}{\mathrm{SO}}
\newcommand{\SU}{\mathrm{SU}}
\newcommand{\ISO}{\mathrm{ISO}}

\begin{document}

\author{Eric Palmerduca}
\email{ep11@princeton.edu}
\affiliation{Department of Astrophysical Sciences, Princeton University, Princeton, New Jersey 08544}
\affiliation{Plasma Physics Laboratory, Princeton University, Princeton, NJ 08543,
U.S.A}

\author{Hong Qin}
\email{hongqin@princeton.edu}
\affiliation{Department of Astrophysical Sciences, Princeton University, Princeton, New Jersey 08544}
\affiliation{Plasma Physics Laboratory, Princeton University, Princeton, NJ 08543,
U.S.A}

\title{Helicity is a topological invariant of massless particles: $C=-2h$}
\date{\today}

\begin{abstract}
There is an elementary but indispensable relationship between the topology and geometry of massive particles. The geometric spin $s$ is related to the topological dimension of the internal space $V$ by $\dim V = 2s + 1$. This breaks down for massless particles, which are characterized by their helicity $h$, but all have 1D internal spaces. We show that a subtler relation exists between the topological and geometry of massless particles. Wave functions of massless particles are sections of nontrivial line bundles over the lightcone whose topology are completely characterized by their first Chern number $C$. We prove that in general $C = -2h$. In doing so, we also exhibit a method of generating all massless bundle representations via an abelian group structure of massless particles.

\end{abstract}

\maketitle

The presence of nontrivial topology is a unifying feature that ties together a diverse array of important physical discoveries of the last 50 years, including the Aharonov-Bohm effect \cite{Aharonov_Bohm}, the Berry phase \cite{Berry1984}, the quantum Hall effects \cite{Thouless1982, Avron1983,Kohomoto1985}, and the existence of topological insulators \cite{Volkov1985}. More recently, it has becomes apparent that nontrivial topology can emerge more generally in many systems supporting waves, such as fluids \cite{Souslov2017,Delplace2017,Tauber2019,Souslov2019,Perrot2019,Faure2022}, plasmas \cite{Yang2016,Gao2016,Parker2020a,Fu2021,Fu2022,Qin2023,Qin2024plasma}, and even the vacuum \cite{bliokh2015,PalmerducaQin_PT, PalmerducaQin_GT,Dragon2024}. The latter refers to the fact that massless particles with nonzero helicity are topologically nontrivial vacuum excitations. In Wigner's classification \cite{Wigner1939,Weinberg1995}, elementary particles are characterized by their geometry rather than their topology, in particular, they are classified as unitary irreducible representations (UIRs) of the (proper orthochronous) \Poincare group $\poincareGroup$. Here, we explore the relationship between the topology and geometry of massless particles and show that massless particles can equivalently be characterized by their topology, and more specifically, by their (first) Chern number $C$. In doing so, we exhibit an abelian group structure of massless particles with respect to the tensor product of line bundles which leads to a very simple method of generating massless particle representations of all helicities from the photon representations.

There is a simple and well-known relationship between the geometry and topology of massive particles. Conventionally, particles are represented by UIRs of $\poincareGroup$ on the space of $n$-component square-integrable wavefunctions over the 4-momentum space $M$, denoted by $L^2(M,\Comp^n)$ \cite{Wigner1939,Weinberg1995}. Note that here we ignore quantum numbers arising from non-spacetime symmetries, such as color charge. For massive particles with $m>0$, $M$ is a mass hyperboloid. According to Wigner's little group method, these representations are classified by the UIRs of the little group $\SO(3) \seq \poincareGroup$ on the internal space $V \doteq \Comp^n$, which in turn are labeled by the non-negative integer or half-integer spins $s$. The spin is related to $\dim V$ by
\begin{equation}\label{eq:spin_dim_relationship}
    \dim V = 2s + 1. 
\end{equation}
Spin characterizes the geometry of the representation while the dimension of a vector space is a topological invariant. Thus, Eq. (\ref{eq:spin_dim_relationship}) gives the elementary but highly useful relationship between the geometry and topology of massive particles.

The situation is very different for massless particles. The spacetime symmetry is again characterized by a UIR of the little group, but for a massless particle the little group is $\SO(2)$ rather than $\SO(3)$ \cite{Wigner1939,Weinberg1995,PalmerducaQin_PT}. The UIRs are labeled not by their spin, but by their (possibly negative) integer helicities $h$ (assuming all elementary massless particles are bosons, as in the standard model). Thus, helicity describes the geometry of massless particles. Unlike the spin representations of $\SO(3)$, the UIRs of $\SO(2)$ are all one-dimensional, so there is no relationship like Eq. (\ref{eq:spin_dim_relationship}) relating the helicity with the dimension of the internal space. For example, although R and L photons have helicities $h = \pm 1$ and R and L gravitons have $h = \pm 2$, each of these four particles have a single polarization degree of freedom \cite{PalmerducaQin_PT,PalmerducaQin_GT}. What then is the relationship between the geometry and topology of massless particles?

The topology of massless particles is far more subtle than that of massive particles. In fact, this topology is essentially obscured in Wigner's conventional representations. The momentum space of massless particles is the forward lightcone
\begin{align}
    \Lightcone &= \{k^\mu= (\omega, \boldsymbol{k}): k^\mu k_\mu = 0, \omega > 0 \} \\
    &\cong \pspace.
\end{align}
In Wigner's representations \cite{Wigner1939}, massless particles are all represented on the same Hilbert space $L^2(\Lightcone,\Comp)$ because the internal spaces are all one-dimensional. However, unlike the massive representations, when $h\neq 0$ these representations are not smooth, meaning that \Poincare transformations typically map a smooth wavefunction to a nonsmooth wavefunction \cite{Flato1983,PalmerducaQin_PT}. To avoid this problem, one instead considers massless particles as UIRs on $L^2(E)$, the space of $L^2$-integrable sections of a complex Hermitian line bundle $\pi:E\rightarrow \Lightcone$ \cite{Simms1968,Flato1983,Asorey1985,PalmerducaQin_PT}. Since $\Lightcone$ has a hole at the origin (as massless particles have no rest frame), it is non-contractible and thus $E$ can be topologically nontrivial (unlike in the massive case when $M$ is a mass hyperboloid). Such lines bundles are topologically characterized by their (first) Chern number $C$ \cite{McDuff2017,PalmerducaQin_PT}; when $C \neq 0$, these particles are topologically nontrivial. We recently showed that photons \cite{PalmerducaQin_PT} and gravitons \cite{PalmerducaQin_GT} are topologically nontrivial; $R$ and $L$ photons have Chern numbers $\mp 2$ and helicities $\pm 1$, while $R$ and $L$ gravitons have Chern numbers $\mp 4$ and helicities $\pm 2$. In this article, we prove that in general $C = -2 h$, giving the precise relationship between the topology and geometry of massless particles. This relationship is the massless analogue of Eq. (\ref{eq:spin_dim_relationship}). That helicity can be viewed as a topological charge was also recently discovered by Dragon \cite{Dragon2024}, but the explicit relationship $C = -2h$ was not given. Furthermore, Dragon used primarily analytic techniques, while we present simpler arguments based on the abelian structure of massless particles under the tensor product.

We first review the vector bundle formalism for massless particles; see Refs. \cite{Simms1968, PalmerducaQin_PT} for additional details. The UIRs of $\poincareGroup$ on the wavefunctions $L^2(E)$ arise from vector bundle representations on $E$. The notion of a representation of a Lie group $G$ on a vector bundle is the natural generalization of representation of $G$ on a vector space. Let $\pi:E\rightarrow \Lightcone$ be a complex Hermitian vector bundle over the forward lightcone $\Lightcone$ with rank $r$ and Hermitian product $\langle\cdot, \cdot \rangle$. We will assume all vector bundles are smooth and have nonzero rank. The fiber at $k$ is denoted by $E_k$ and a vector $v\in E$ may be written as $(k,v)$ to indicate that $v \in E_k$. $E$ is a bundle representation of a Lie group $G$ if there are smooth $G$-actions on $\Lightcone$ and $E$,
\begin{align}
    k &\mapsto gk \\
    (k,v) &\mapsto (gk,gv),\end{align}
such that $g$ linearly maps $E_k$ onto $E_{gk}$. In particular, the action commutes with projection: $\pi g(k,v) = gk$. The representation is unitary if $g\in G$ maps $E_k$ to $E_{gk}$ unitarily, that is, if
\begin{equation}
    \langle g v_1, g v_2\rangle = \langle v_1 , v_2 \rangle
\end{equation}
for any $v_1,v_2$ in the same fiber. A subrepresentation of $E$ is a subbundle $F$ of $E$ such that $F$ is a representation of $G$ under the same $G$-action. $E$ is irreducible if it has no proper subbundles. Two bundle representations $E_1$ and $E_2$ of $G$ are unitarily equivalent if there is a unitary vector bundle isomorphism $\psi:E_1 \rightarrow E_2$ which is equivariant with respect to the group action:
\begin{equation}
    \psi(g v) = g\psi(v)
\end{equation}
for any $g \in G$ and $v \in E_1$. Note in particular that this implies $E_1$ and $E_2$ have the same topology as vector bundles.

We will primarily be concerned with the case when $G=\poincareGroup = \mathbb{R}^{3+1} \rtimes \SO^+(3,1)$, where $\rtimes$ denotes the semidirect product and $\SO^+(3,1)$ the connected component of the Lorentz group. A generic element of $\poincareGroup$ can be written as $T_a \circ \Lambda$ where $T_a$ is a translation by $a \in \mathbb{R}^{3+1}$ and $\Lambda \in \SO^+(3,1)$. We assume that $\SO^+(3,1)$ acts on $\Lightcone$ by the standard 4-vector action $k \mapsto \Lambda k$.

Massless particles can be considered as unitary irreducible UIRs of $\poincareGroup$ on vector bundles over $\Lightcone$ in which translations $T_a$ act via \cite{Simms1968,PalmerducaQin_PT}
\begin{equation}\label{eq:momentum_condition}
    T_a(k,v) = e^{ik^\mu a_\mu}(k,v).
\end{equation}
We will call such a representation a massless bundle UIR. Condition (\ref{eq:momentum_condition}) ensures that $(k,v)$ has momentum $k$. As in the case of Wigner's representations \cite{Wigner1939, Weinberg1995}, the massless bundle UIRs of $\poincareGroup$ are classified by their actions of the little group, which are in turn labeled by the integer helicities $h$ \cite{PalmerducaQin_PT}. In particular, every such bundle $E$ is a line bundle ($r=1$) such that every vector in $E$ is an eigenvector of the helicity operator $\boldsymbol{\hat{k}}\cdot \boldsymbol{J}$ with eigenvalue $h$:
\begin{equation}\label{eq:helcity_action}
    (\boldsymbol{\hat{k}}\cdot \boldsymbol{J}) v = hv
\end{equation}
Here, $\boldsymbol{J}$ is the generator of the $\SO(3) \seq \poincareGroup$ action on $E$, meaning that if $R_{\hat{\boldsymbol{k}}}(\theta)$ is a rotation by $\theta$ about $\hat{\boldsymbol{k}}$ then
\begin{equation}
    R_{\hat{\boldsymbol{k}}}(\theta) = e^{-i\hat{\boldsymbol{k}} \cdot \boldsymbol{J}}.
\end{equation}
Eq. (\ref{eq:helcity_action}) means that 
\begin{equation}
    R_{\hat{\boldsymbol{k}}}(\theta)(k,v) = e^{-i\boldsymbol{\hat{k}}\cdot \boldsymbol{J}\theta} (k,v) = (k,e^{-ih\theta}v).
\end{equation}
Since $E$ is a bundle representation of $\poincareGroup$, it is necessary that a $\theta = 2\pi$ rotation acts trivially, so $h$ must be an integer. Note that if one considers the generalization to projective bundle representations, then half-integer helicities can also be obtained. Since we are assuming all massless particles are bosons, we can ignore the projective case. Two massless bundle UIRs are unitarily equivalent if and only if they have the same helicity (\cite{PalmerducaQin_PT}, Thm. 35). In this way, $h$ fully characterizes the geometry of the these representations. However, a priori, it is not obvious that particles with different helicities have different topologies. We will prove that this is indeed the case.

The examples of the R and L photon bundles are exceptionally important, as we will show that all other massless particles can be constructed from them. We will review the photon bundles here; additional details can be found in Ref. \cite{PalmerducaQin_PT}. The total photon bundle $\gamma_T$ is defined as the collection of all eigenmodes (photons) in Fourier space of the vacuum Maxwell's equations, which form a vector bundle over $\Lightcone$. In particular, the solutions are labeled by the Fourier transformed electric field $\boldsymbol{E}$, so the fiber at $k^\mu = (|\boldsymbol{k}|,\boldsymbol{k})$ is the 2D complex vector space defined by $\{\boldsymbol{E} \in \Comp^3|\boldsymbol{k}\cdot \boldsymbol{E} = 0\}$. It is a massless unitary representation of $\poincareGroup$ defined by the standard Lorentz transformations of $\Lightcone$ and the electric field. That is, $\boldsymbol{E}$ transforms under rotations like a 3-vector, under translations $T_a$ by $\boldsymbol{E} \rightarrow \boldsymbol{E}e^{ik^\mu a_\mu}$, and under boosts by the well-known relativistic electromagnetic boost transformation \cite{Jackson1999,PalmerducaQin_PT}. $\gamma_T$ is, however, a reducible bundle representation:
\begin{equation}
    \gamma_T = \gamma_1 \oplus \gamma_{-1}
\end{equation}
where $\gamma_{\pm 1}$ are the line subbundles consisting of $R$- and $L$-polarized photons, respectively. More explicitly, $\gamma_{\pm 1}$ contains all vectors of the form $(k,\boldsymbol{E})$ where $\boldsymbol{E} = \alpha(\boldsymbol{e}_1 \pm i\boldsymbol{e}_2)$, $\alpha \in \Comp$, and $(\boldsymbol{e}_1,\boldsymbol{e}_2,\boldsymbol{k})$ is a right-handed coordinated system of $\mathbb{R}^3$. That $\gamma_{\pm 1}$ are also representations of $\poincareGroup$ reflects the fact that $R$- and $L$-polarizations are preserved under \Poincare transformations. These bundles have helicity $h=\pm 1$ and Chern number $C = \mp 2$. The former comes from
\begin{align}
    R_{\boldsymbol{\hat{k}}}(\theta) (k,\boldsymbol{E}_\pm) &= e^{-i (\boldsymbol{\hat{k}} \cdot \boldsymbol{J})\theta }(k,\boldsymbol{E}_\pm) \\
    &= (k,e^{\mp i\theta}\boldsymbol{E}_\pm),
\end{align}
and the latter can be explicitly computed from the Berry connection or the clutching construction \cite{PalmerducaQin_PT}.

We present a construction of massless bundle UIRs of arbitrary helicity which is motivated by observations about the little group representations. In both Wigner's formalism \cite{Wigner1939,Weinberg1995} and the vector bundle formalism \cite{Simms1968,PalmerducaQin_PT,PalmerducaQin_GT}, irreducible representations of $\poincareGroup$ are related to representations of the little group. The little group $H_k$ at a fixed $k$ in the momentum space $M$ is defined as the subset of $\lorentzGroup$ which leaves $k$ invariant. The little groups for different $k$ are related by $H_{\Lambda k}=\Lambda H_{k}\Lambda^{-1}$ for $\Lambda \in \lorentzGroup$. Since any two $k$ in $M$ are related by some Lorentz transformation, the isomorphism type of the group $H_k$ is independent of $k$; the isomorphism type of $H_k$ is called the little group $H$ of the representation. For massive particles $M$ is a mass hyperboloid, and $H = \SO(3)$. However, for massless particles with $M=\Lightcone$, $H=\ISO(2)=\mathbb{R}^2 \rtimes \SO(2)$, the group of (inhomogeneous) isometries of $\mathbb{R}^2$ \cite{Wigner1939}. Representations in which the translations (the $\mathbb{R}^2$ part of $\ISO(2)$) act nontrivially on $E$ correspond to representations with a continuum of degrees of freedom. Since there are no known particles exhibiting such continuous degrees of freedom, we make the conventional assumption that $\mathbb{R}^2$ translations act by the identity on $E$ \cite{Weinberg1995, Maggiore2005}. Thus, the little group essentially acts like $\SO(2)$. The finite dimensional vector space UIRs of $\SO(3)$ and $\SO(2)$ are the well-known spin and helicity representations, respectively. The spin representations of $\SO(3)$ are labeled by the non-negative integers $s$, and have complex dimension $2s+1$. In contrast, the helicity representations are labeled by the complete set of integers $h$, but always have complex dimension $1$. If $R(\theta) \in \SO(2)$ is a rotation by angle $\theta$, then in the helicity $h$ representation $R(\theta) v = e^{-ih\theta}v$ for any vector $v$. One can ask the following question: given two UIRs, can one produce another UIR? Suppose $U_1$ and $U_2$ are two UIR of a group $G$ on complex vector spaces $V_1 \cong \Comp^{n_1}$ and $V_2\cong \Comp^{n_2}$. The direct sum $U_1 \oplus U_2$ and direct product $U_1 \otimes U_2$ are well-defined representations of $G$ on $V_1 \oplus V_2$ and $V_1 \otimes V_2$ with dimensions $n_1 + n_2$ and $n_1n_2$, respectively \cite{Hall2015}. For $v_i \in V_i$, they are defined by
\begin{gather}
    g(v_1+v_2) = gv_1 + gv_2 \\
    g(v_1 \otimes v_2) = gv_1 \otimes gv_2.
\end{gather}
These representations are unitary under the Hermitian products
\begin{gather}
    \langle (v_1+v_2),(w_1+w_2) \rangle  = \langle v_1, w_1\rangle + \langle v_2, w_2\rangle \\
    \langle v_1\otimes v_2, w_1 \otimes w_2\rangle = \langle v_1, w_1\rangle \langle v_2, w_2\rangle \label{eq:Hermitian_tensor_product}
\end{gather}
where the Hermitian products on the rhs are those on $V_1$ and $V_2$. 
$U_1 \oplus U_2$ is obviously reducible. Typically, $U_1 \otimes U_2$ is also reducible; the problem of decomposing $U_1 \otimes U_2$ into UIRs is the Clebsch-Gordon problem \cite{Hall2015}. The most well-known cases are those of $\SO(3)$ and $\SU(2)$, which satisfy:
\begin{equation}\label{eq:spin_addition}
    S_l \otimes S_m \cong S_{l-m} \oplus S_{l-m + 1} \oplus \ldots \oplus S_{l+m},
\end{equation}
where $S_k$ is the spin $k$ representation and $l \geq m$. In particular,
\begin{equation}\label{eq:spin_addition_S_1}
    S_1 \otimes S_1 \cong S_0 \oplus S_1 \oplus S_2.
\end{equation}
However, if $U_1$ and $U_2$ both have dimension $1$, then $U_1 \otimes U_2$ also has dimension $1$ and is therefore irreducible. Thus if $W_h$ is the helicity $h$ representation, then $W_{h_1} \otimes W_{h_2}$ is irreducible. In fact,
\begin{align}
    R(\theta)(v_1\otimes v_2) &= e^{-ih_1\theta}v_1 \otimes e^{-ih_2 \theta}v_2 \\
    &= e^{-i(h_1+h_2)\theta}(v_1 \otimes v_2),
\end{align}
so
\begin{equation}
    W_{h_1} \otimes W_{h_2} \cong W_{h_1+h_2}.
\end{equation}
When $h$ is positive,
\begin{equation}
    W_h \cong \bigotimes_{i=1}^h W_1 \doteq W_1^h.
\end{equation}
Since $W_1 \otimes W_{-1} \cong W_0$, we can define $W_1^{-1} = W_{-1}$, and then 
\begin{equation}
    W_h \cong W_1^h
\end{equation}
for any $h \in \mathbb{Z}$. Thus, the helicity representations are not just in bijective correspondence with $\mathbb{Z}$, but they are actually isomorphic to $\mathbb{Z}$ as an abelian group, with the tensor product as the group operation. As such, they are all generated by $W_1$ (or $W_{-1})$. This is a distinctly massless phenomena, as Eq. (\ref{eq:spin_addition_S_1}) shows that a similar result does not apply to the massive spin representations. Indeed, the spin representations are labeled by non-negative integers, which do not even form a group.

We now show that we can construct all the bundle UIRs of massless particles in a similar way, by taking tensor products of $\gamma_{\pm 1}$. Suppose we have two massless bundle UIRs $E_1$ and $E_2$ with helicities $h_1$ and $h_2$. Let $E_1 \otimes E_2$ be the tensor product of the bundles, which is a line bundle over $\Lightcone$ in which the fiber at $k$ is the vector space $E_{1,k} \otimes E_{2,k}$ \cite{Tu2017differential}. It carries a Hermitian product given on each fiber by Eq. (\ref{eq:Hermitian_tensor_product}). We can define a unitary representation of $\poincareGroup$ on $E_1 \otimes E_2$ as follows:
\begin{gather}
    \Lambda(k,v_1 \otimes v_2) = (k, \Lambda v_1 \otimes \Lambda v_2) \label{eq:Lorentz_action}\\
    T_a (k, v_1 \otimes v_2) = e^{ik^\mu a_\mu}(k,v_1 \otimes v_2) \\
    (T_a \circ \Lambda)(k,v_1\otimes v_2) = T_a [\Lambda(k,v_1 \otimes v_2)].
\end{gather}
To verify that this is actually a representation of $\poincareGroup$, it is necessary to check that
\begin{equation}
    (\Lambda \circ T_a)(k,v_1 \otimes v_2) = \Lambda[T_a (k,v_1 \otimes v_2)],
\end{equation}
which follows immediately from the fact that $$\Lambda \circ T_a \circ \Lambda^{-1} = T_{\Lambda a}.$$
Since $E_1 \otimes E_2$ has rank 1, this representation is irreducible. Moreover, by Equation (\ref{eq:Lorentz_action})
\begin{equation}
    e^{i(\boldsymbol{\hat{k}} \cdot \boldsymbol{J})\theta}(k,v_1\otimes v_2) = e^{i(h_1+h_2)\theta}(k,v_1 \otimes v_2),
\end{equation}
showing that $E_1 \otimes E_2$ has helicity $h_1+h_2$. Defining $\gamma_{1}^{-1} \doteq \gamma_{-1}$ and $\gamma_h \doteq \gamma_1^h \doteq \bigotimes_{i=1}^h \gamma_1^{\text{sgn}(h)}$, we have that $\gamma_h$ is a massless bundle UIR with helicity $h$. In this way we obtain smooth bundle representations for massless particles with any integer helicity. By Theorem 35 of Ref. \cite{PalmerducaQin_PT}, every massless particle is unitarily equivalent to one of these representations.

Determining the topology of $\gamma_h$ is also a simple matter. Recall that the topology of a line bundle over $\Lightcone$ is completely determined by its Chern number $C$. The Chern number can be considered as a measure of how ``twisted'' or nontrivial a vector bundle is; in particular, $C=0$ corresponds to the trivial bundle \cite{Bott2013}. For line bundles, the Chern number is additive with respect to the tensor product (Ref. \cite{Bott2013}, Equation 20.1)
\begin{equation}
    C(E_1 \otimes E_2) = C(E_1) + C(E_2).
\end{equation}
Note that this is a special property of line bundles which does not generally hold for higher rank bundles. As the R and L photon bundles have Chern numbers $C(\gamma_{\pm 1}) = -2$, we obtain
\begin{equation}
    C(\gamma_h) = C(\gamma_1^h) = -2h.
\end{equation}
This equation describes the fundamental relationship between the geometry and topology of massless particles. One consequence is that the helicity can be considered as a topological invariant, despite being defined purely in terms of \Poincare geometry. Another consequence is that massless particles with different helicity are topologically distinct, with the helicity as a measure of how globally twisted the particle space is. This is not readily detectable using the (nonsmooth) Wigner representations of massless particles \cite{Wigner1939,Weinberg1995}, in which the space of wavefunctions of all massless particles are represented on the same Hilbert space $L^2(\Lightcone,\Comp)$. In deriving this relationship, we have illustrated a particularly simple method of constructing smooth bundle representations for any massless particle, namely, by taking tensor products of the photon bundles $\gamma_{\pm 1}$. This is to be compared with the more complicated method of Asorey et al. \cite{Asorey1985} which involves taking the quotient of two tensor product spaces, or equivalently, by applying a series of gauge conditions to symmetric traceless rank-$h$ tensors.

As an example, we can immediately construct the R and L graviton bundles. These bundles are given by $\Xi_\pm \doteq \gamma_{\pm1} \otimes \gamma_{\pm1}$ with the plus and minus signs corresponding to the R and L graviton bundles, respectively. At $k = (|\boldsymbol{k}|,\boldsymbol{k})$, choose any right-handed basis $(\boldsymbol{e}_1,\boldsymbol{e}_2,\boldsymbol{k})$ of $\mathbb{R}^3$. The photon fiber $\gamma_\pm(k)$ is spanned by $v_\pm = \boldsymbol{e}_1 \pm i \boldsymbol{e}_2$, so the graviton fiber $\Xi_\pm(\boldsymbol{k})$ is spanned by $v_\pm \otimes v_\pm$. By expressing the tensor product as an outer product, \begin{equation}
    (v_\pm \otimes v_\pm)^{ij} = v_\pm^i v_\pm^j,
\end{equation}
$\Xi_{\pm}(k)$ is embedded in the vector space of three-by-three symmetric tensors. Furthermore, since $(v_\pm \otimes v_\pm) \cdot \boldsymbol{k} = 0$ and $\sum_{i=1}^3 v_\pm^i v_\pm^i = 0$,  the vectors of $\Xi_\pm (k)$ are automatically in the typical transverse-traceless gauge commonly used to describe gravitons \cite{Maggiore2005,Schutz2009,PalmerducaQin_GT}. Notice that in this construction it was never necessary to manually impose the transverse-traceless gauge conditions. Whichever gauge fixing one chooses for the photon bundle automatically propagates through to all higher helicity bundles. A comparison with Ref. \cite{PalmerducaQin_GT} shows that the tensor product method gives an equivalent and simpler construction of the $R$ and $L$ graviton bundles.

It has long been known that the limit of zero particle mass is singular \cite{Wigner1939}. This is conventionally viewed as a geometric singularity, as the little group jumps abruptly from being $\SO(3)$ for massive particles to $\SO(2)$ for massless particles. However, there is an accompanying topological singularity that occurs too, as the momentum space jumps from being a contractible mass hyperboloid to the non-contractible lightcone $\Lightcone$. This allows for massless particles to have nontrivial topology, as illustrated in the bundle representations. We have shown that this singular limit involves other non-obvious differences between massive and massless particles, such as a completely different relationship between the topology and geometry as well as the emergence of an abelian group structure of massless particles. The topological analysis of elementary particles is still fairly young, and we suspect other important implications of the topological nontriviality of massless particles will be discovered.

\begin{acknowledgments}
This work is supported by U.S. Department of Energy (DE-AC02-09CH11466).
\end{acknowledgments}


\bibliography{helicity}
\end{document}